\documentclass[letterpaper]{article} 
\usepackage{aaai2027}
\nocopyright
\usepackage[hyphens]{url}  
\usepackage{graphicx} 
\usepackage{natbib}  
\usepackage{caption} 
\usepackage{algorithm}
\usepackage{algorithmic}

\usepackage{booktabs}

\usepackage{amsfonts}
\usepackage{amsmath, amssymb, amsthm}
\usepackage{multirow}

\title{Training Small LLMs as Spatial Multi-Agent Policies}

\author{
    Yi Mao,
    Andrew Perrault
}
\affiliations{
    Department of Computer Science and Engineering, The Ohio State University\\
    \texttt{mao.496@osu.edu}, \texttt{perrault.17@osu.edu}
}

\begin{document}

\maketitle
\begin{abstract}

Training LLM-based multi-agent systems with multi-agent reinforcement learning is rapidly gaining traction, and a parallel line of work argues that such systems should be judged by their behavior, not only their reward. We take up both threads in spatial cooperative games, where small frozen LLMs prompted with low-level actions fail outright, earning zero reward. Guided by the options/semi-MDP framework---and, because option execution is asynchronous across agents, its multi-agent extension in macro-action Dec-POMDPs---we equip each game with a library of symbolic \emph{options}: typed, state-feasible, short-horizon behaviors executed by a symbolic planner. Each library is drafted by a frontier coding model from the game's source code; the feasibility guards that filter each menu are then synthesized mechanically from cheap random-policy burn-in rollouts---a guard is adopted only if it explains repeated execution failures while hiding no logged success---so no guard is authored, selected, or reward-tuned by hand. Each agent's LLM acts as its policy over options, with a private per-agent LoRA adapter trained by a per-agent variant of multi-agent GRPO (PA-MAGRPO); this lifts frozen bases from zero reward to competent play across three games and four small backbones. Behavioral audits then reveal that reward and cooperation decouple: a rising reward curve may simply mean that one agent has learned to run the entire task alone while its partner idles---cooperation emerges only when the task makes it necessary. Reward alone is thus an unreliable readout of cooperation; behavioral evaluation must sit alongside it.

\end{abstract}
\section{Introduction}
\label{sec:intro}

Training LLM-based multi-agent systems with multi-agent reinforcement
learning is rapidly gaining traction: recent work moves beyond prompting frozen
frontier models~\citep{mosquera2024meltingpot, gallego2026cooperation}
and optimizes the interacting agents
themselves~\citep{Cory, liu2025magrpo, piche2025robust}, while a
parallel line of work argues that such systems should be judged by their
behavior rather than by reward alone~\citep{leibo2017ssd}. A frozen model can only be
steered by prompting, so its coordination failures are permanent; a
trainable policy can instead turn environment feedback into durable,
agent-specific behavior. But joint RL multiplies
rollout and gradient cost by the number of agents, pushing the trainable
model size to small models (2--4B) under our academic budget. This paper asks what it
takes to train small LLMs as \emph{spatial} multi-agent policies---and what
actually emerges when training succeeds.

Three obstacles stand in the way. First, a \emph{grounding} gap: in
our experiments, small frozen models dropped into a 2D spatial game could not
reliably turn a high-level goal (``clean the dirt to my north'') into the
primitive actions the environment expects---they hallucinate spatial
preconditions and collapse onto a single ``safe''
action~\citep{valmeekam2023planning, ahn2022saycan, huang2022innermonologue,
liang2023codeaspolicies}---so naive fine-tuning starts from, and never
leaves, zero reward. Second, a \emph{horizon} problem: episodes run for
hundreds of primitive steps under a sparse team reward, leaving the optimizer
almost no per-decision credit signal. Third, a \emph{specialization} problem:
cooperation often requires agents to behave differently, but a team trained
through one shared set of weights averages conflicting role gradients and
collapses onto identical behavior. A workable
recipe has to address all three.

The first two obstacles have a classic remedy: temporal abstraction.
Instead of asking the LLM to steer with primitive commands (up, down,
left, right, interact), we let it choose among a small set of
\emph{options}~\citep{sutton1999options}. An option is a named,
short-horizon behavior with a target---$\textsc{move\_to}(x,y)$,
$\textsc{pick\_up}(\text{onion})$, $\textsc{clean}(\text{nearest
dirt})$. Once the agent selects $\textsc{move\_to}(x,y)$, a symbolic
executor computes a shortest-path route and emits the primitive moves:
one decision spans multiple environment steps. Each option
also carries checking code that tests, against the current state and
observation, whether the option makes sense right now---and an option that
fails its checks is simply not shown. If no apple is in the agent's local
view, \textsc{eat} does not appear in the prompt at all. At a decision
point the LLM thus sees a short lettered menu of feasible options and
simply picks one: it decides \emph{what} to do, never \emph{how}.
Decisions become far fewer and semantically meaningful. Formally, each agent now faces a semi-MDP, and
because options start and finish asynchronously across agents, the
team-level problem is a macro-action Dec-POMDP~\citep{amato2014macdec}, in
whose terms our method is decentralized macro-level policy
learning~\citep{xiao2020macro} with LLM policies.

Each game's option library is built with exactly one frontier-model step:
Claude Opus 4.8 reads the game's source code and drafts a typed option
schema in which every option carries an \emph{effect predicate}---a
post-condition stating what the option promises to do. The feasibility
guards that filter the menu are then synthesized by a fixed algorithm, with
no model and no human in the loop: cheap uniform-random burn-in rollouts
label every execution as valid or vacuous under the effect predicate, and
a guard is adopted only if it explains repeated failures while hiding zero
logged successes---a criterion of execution validity, never reward. After
construction the frontier model is not queried again---not during
training, evaluation, or action selection. The specialization
obstacle motivates the second ingredient: a private LoRA
adapter~\citep{LoraHu2021} per agent on a shared frozen base, trained
jointly with a per-agent variant of multi-agent
GRPO~\citep{liu2025magrpo, grpo} (\emph{PA-MAGRPO}): the shared base
retains common language and task knowledge, while private adapters give
divergent roles a parameter slot in which to emerge. Neither ingredient is
redundant: the frozen base with the same option menus stays near zero reward
on two of three games, and a single shared adapter collapses toward identical
behavior.

The recipe works: it lifts frozen bases from zero reward to competent
play on all three games, and does so across four small backbones. We can also say \emph{why}. Option-level decisions
compress the horizon: the LLM is consulted on only a fraction of
environment steps. The action space is semantic, where a language
model's priors help. And feasibility filtering removes the grounding
failures outright. Training success makes the question
visible: holding the recipe fixed, does a rising reward curve mean the
team cooperates?

Behavioral audits show the answer is: not necessarily. Across our
three games, the same recipe yields high reward with genuine division of
labor, high reward with one agent working while its partner idles, and
nearly flat reward that hides a real shift to sustainable harvesting;
coordination emerges only where the task demands it. Cooperation tracks
the task's incentive structure, not the reward curve.

\paragraph{Contributions.}
\begin{enumerate}
\item 
\textbf{A symbolic option library that makes small
LLMs viable spatial MARL policies.} Each agent acts as a policy over a fixed
library of symbolic options---the \emph{option menu} interface: the LLM
chooses \emph{what} to do from a state-filtered menu, and a symbolic planner
handles \emph{how}. Formally, the interface
recasts each agent's problem as a semi-MDP over options and the team's as a
macro-action Dec-POMDP~\citep{sutton1999options, amato2014macdec}.
Without it, neither the frozen base nor the fine-tuned policy produces
meaningful behavior on the harder games.

\item \textbf{Per-agent reinforcement fine-tuning (RFT) for multi-agent policies.} We pair the option menu
with per-agent LoRA adapters trained by MAGRPO (\emph{PA-MAGRPO}). The
shared base preserves common language and task knowledge, while private adapters
give each agent a parameter slot where divergent behavior can emerge under a
shared team return.  This directly addresses the specialization
obstacle above: it avoids the gradient-averaging failure of a single shared
adapter and enables cleaner/eater specialization
in Cleanup without explicit role labels.

\item \textbf{Evidence that reward and cooperation decouple.} Reward
and cooperation decouple in both directions: Overcooked Asymmetric
Advantages yields high reward from a one-worker/one-idler team (with Forced
Coordination as the within-game control where deliveries do certify
hand-offs), while Commons Harvest improves patch survival substantially
under a nearly flat reward---reward curves can both overstate and
understate cooperation.
\end{enumerate}

\section{Related Work}
\label{sec:related}

\paragraph{Training LLM multi-agent systems.}
A growing line of work \emph{trains} interacting LLM agents rather
than prompting them:
CORY~\citep{Cory} coevolves two copies of an LLM on text tasks;
MAGRPO~\citep{liu2025magrpo} introduces a group-relative multi-agent
objective for collaborative writing and coding; Piche et
al.~\citep{piche2025robust} fine-tune open LLMs toward non-exploitable
cooperation in textual social dilemmas; YOLO-MARL~\citep{YoloMarl} keeps
the LLM outside the gradient loop entirely, querying it once to synthesize
a planner. We adopt MAGRPO's group-relative objective but move it into
embodied, partially observable spatial games, with per-agent
adapters---a design established by LoRASA~\citep{zhang2025lorasa} on SMAC
and MAMuJoCo and by multi-LoRA coding pipelines~\citep{lee2026mapcoderlite}.
A second line argues that team reward alone is an insufficient signal
of cooperation: sequential social dilemmas~\citep{leibo2017ssd} and
common-pool resource models~\citep{perolat2017commonpool} measure behavior
directly through social-outcome metrics (efficiency, equality,
sustainability, peace); inequity aversion~\citep{hughes2018inequity}
improves cooperation without altering the task reward; Biswas et
al.~\citep{biswas2025interdependence} show that MARL agents on Overcooked
can attain high reward with very low \emph{interdependence}; and
Shapley-Coop~\citep{hua2025shapleycoop} finds that self-interested LLM
agents default to non-cooperation absent explicit credit mechanisms. These
works establish, in single environments or single directions, that reward
and cooperation can diverge; we join the two lines by training the
agents ourselves and extending the divergence to a \emph{bidirectional,
cross-task} decoupling under one fixed recipe.

\paragraph{Temporal abstraction and macro-action MARL.}
An option couples an initiation set, an intra-option policy, and a
termination condition, and acting over a fixed option set turns an MDP
into a semi-MDP~\citep{sutton1999options}; with options terminating
asynchronously across agents, the planning formalism is the macro-action
Dec-POMDP~\citep{amato2014macdec}, over which macro-action-based deep MARL
trains decentralized policies~\citep{xiao2020macro}. Our option menu
instantiates this design with a frontier-drafted library where prior work
typically hand-specifies one, and with a fine-tuned LLM rather than a
task-specific network as the policy over options; unlike option-discovery
methods, our options are fixed and symbolic, and learning happens only in
the policy over them.

\paragraph{Closest neighbors.}

Our design builds directly on three threads. \citet{mosquera2024meltingpot} show that Melting Pot cooperation
can be posed to LLMs through a compound-action vocabulary
(\texttt{go\_to(x,y)}, \texttt{immobilize}, \texttt{stay\_put},
\texttt{explore}); we adopt essentially their vocabulary and ask what changes
when the model can \emph{learn}, rather than act as a frozen GPT-3.5/4
reasoner. \citet{gallego2026cooperation} shows that a frozen
frontier model can author Python policies for social dilemmas under
social-metric feedback; we likewise use a frontier model as a code author, but
only at construction time---learning then happens inside the agents. \citet{piche2025robust} demonstrate group-relative fine-tuning
toward cooperation in textual social dilemmas such as Trust-and-Split; we
bring that objective family onto spatial substrates with macro actions.
Combining the three---spatial games, per-step small-LLM policies, and
gradient-based multi-agent training---is to our knowledge new; the
combination lets us hold one training recipe fixed across tasks and compare
the behavior that emerges. Because these works differ in agent counts, horizons, and evaluation
metrics, no direct numerical comparison is possible; we instead benchmark
against frozen-base and ablation variants under an identical action
interface, reporting behavioral metrics alongside return.


\section{Background}
\label{sec:background}

\paragraph{From Dec-POMDPs to macro-action Dec-POMDPs.}
The underlying game is a cooperative decentralized POMDP
(Dec-POMDP)~\citep{amato2014macdec} given by
$(\mathcal{S}, \{\mathcal{O}^{i}\}_{i=1}^{N}, \{\mathcal{A}^{i}\}_{i=1}^{N},
\Omega, P, R, \gamma)$: at each step the environment is in state
$s_t \in \mathcal{S}$, each agent $i$ receives a private observation
$o^i_t = \Omega(s_t, i)$, the joint action
$\mathbf{a}_t = (a^1_t, \dots, a^N_t)$ drives the transition
$P(s_{t+1} \mid s_t, \mathbf{a}_t)$, and all agents share the team reward
$r_t = R(s_t, \mathbf{a}_t)$. The objective is a policy set
$\boldsymbol{\pi} = \{\pi^{i}\}_{i=1}^{N}$ maximizing
$J(\boldsymbol{\pi}) = \mathbb{E}[\sum_{t} \gamma^{t} r_t]$.

Acting at the primitive level is not the only choice. An \emph{option}
$z = \langle \mathcal{I}_z, \pi_z, \beta_z \rangle$ couples an initiation set
$\mathcal{I}_z$ (the observations in which $z$ may be selected), an
intra-option policy $\pi_z$ (here, a symbolic planner emitting primitive
actions from $\mathcal{A}^{i}$), and a termination condition $\beta_z$ (here,
completion, failure, or timeout of the macro)~\citep{sutton1999options}.
Acting over a fixed option set $\mathcal{Z}^{i}$ turns agent $i$'s decision
problem into a \emph{semi-MDP}: decisions occur only at option boundaries,
separated by a variable number of environment steps. We write
$\mathcal{Z}^{i}(o^i_t) = \{z \in \mathcal{Z}^{i} : o^i_t \in \mathcal{I}_z\}$
for the \emph{menu} of options available to agent $i$ at observation $o^i_t$;
a \emph{policy over options} $\pi^{i}(z \mid o^i_t)$, supported on
$\mathcal{Z}^{i}(o^i_t)$, selects a new option whenever the previous one
terminates.
Because options begin and end asynchronously across agents, the team-level
process is not itself a single semi-MDP; the resulting model---the Dec-POMDP
above with each $\mathcal{A}^{i}$ replaced by the option library
$\mathcal{Z}^{i}$---is a \emph{macro-action
Dec-POMDP}~\citep{amato2014macdec, xiao2020macro}: whenever agent $i$'s
current option terminates, it selects a new option from its menu while its
teammates continue executing theirs.

\paragraph{Group-relative policy optimization.}
GRPO~\citep{grpo} replaces the learned value baseline of PPO-style methods
with a cohort statistic: it samples $K$ rollouts, normalizes their returns
within the cohort, and uses the normalized return as the advantage in a
clipped surrogate update with a KL penalty toward a reference policy.
MAGRPO~\citep{liu2025magrpo} extends this to cooperative multi-agent
training: rollouts are joint, the scored quantity is the shared team
return, and every agent's update uses the same cohort-normalized
advantage---avoiding a centralized critic, which is expensive and unstable
at LLM scale. The Method section gives our per-agent instantiation.
\section{Two Failure Modes of a Small LLM as a Cooperative Spatial Agent}
\label{sec:method-failures}

Before describing our recipe, we identify two failure modes of a na\"ive deployment on Cleanup: the frozen Qwen3-4B-Instruct-2507 base, no fine-tuning, is handed the standard prompt and asked to emit a primitive action letter \texttt{A}--\texttt{I} (movement, turning, \textsc{fire\_zap}, \textsc{fire\_clean}). Both failures are visible directly in the input/output stream.

\paragraph{Failure 1 --- spatial hallucination.}
The prompt tells the LLM its position, facing direction, nearest dirt, and a layout hint---everything needed to reason \emph{``to fire at dirt to my north, I must first face north.''} In practice it does not. A verbatim trace from the first step of an episode:

\begin{quote}\scriptsize
\textbf{Prompt to agent 0} (truncated):\\
\texttt{You are agent 0. World position: (14,9). Facing: West.}\\
\texttt{Sensors: Dirt visible 13, nearest (13,5) 5st, (15,5) 5st.}\\
\texttt{Layout: SAND. River is to your north (rows 1--5).}\\
\textbf{Response from agent 0:}\\
\texttt{state: river\_dirty} \\
\texttt{choice: I~~~~~~~~~~~~~~~~~~~~~~~~~$\leftarrow$~FIRE\_CLEAN}
\end{quote}

The agent faces \emph{west}; the only visible dirt is \emph{north} of it; the beam discharges harmlessly into open sand. Across all five evaluation episodes ($7$ agents $\times$ $150$ steps), the LLM never once precedes a \textsc{fire\_clean} with the \textsc{turn\_left}/\textsc{turn\_right} actions that the prompt itself enumerates as legal.  The model has the symbols and ignores their meaning.  We call this \emph{spatial hallucination}: the LLM produces plausible-looking decisions whose pre-conditions (here, ``facing north'') it has not verified.

\paragraph{Failure 2 --- mode collapse on a ``safe'' default.}
Even when no dirt is in view, the LLM emits the same letter: agents seeing \texttt{Dirt visible: 0} still respond \texttt{choice: I}. Aggregated over all 5{,}250 LLM calls, the action distribution collapses to a single letter: \textsc{fire\_clean} $95.8\%$, \textsc{noop} $4.2\%$, and the other seven primitives---including the turning actions that would let cleaning succeed---exactly $0\%$.

Worse, applying single-agent RL fine-tuning on the same prompt format does not unstick this attractor; it \emph{sharpens} it: \textsc{fire\_clean} share rises from $95.8\%$ to $98.4\%$ (Ablation~A, \S\ref{sec:exp}).  The cleaning bias lives in the base model's prior over single-letter responses to a \texttt{river\_dirty} prompt, and per-step letter-level RL only reinforces it.  We call this \emph{mode collapse on a safe default}: when the action interface forces a one-token-at-a-time micro-policy, our tested model picks the most semantically-loaded letter --- the one whose macro-name evokes ``cleaning'' --- and refuses to leave it.

The two ingredients of our method target these two failures one-for-one;
we describe them next and verify empirically that they fix the failures in
the Experiments.

\section{Method}
\label{sec:method}

\subsection{A Symbolic Option Library and Its Menu Interface}
\label{sec:method-ip}

We lift the LLM's decision space from primitive actions to the fixed
library of symbolic options of \S\ref{sec:background}, presented at each
decision point as a lettered \emph{menu}. The mapping onto the option
formalism is exact: deterministic feasibility checks implement the
initiation set $\mathcal{I}_z$; a symbolic executor implements the
intra-option policy $\pi_z$, expanding a chosen option into primitive
actions through shortest-path movement, orientation changes, and
interaction preconditions; and completion, failure, or timeout of the
macro implements termination $\beta_z$. Feasibility is decided by code,
not by a model---path reachability on the current map, held-item
preconditions (placing an onion requires holding one), and task-state
gates (plating requires a ready pot)---and an option that fails any check
is simply not shown. The LLM, prompted with its local text observation and
the surviving lettered menu, samples one option
$z^{i}_{t} \in \mathcal{Z}^{i}(o^{i}_{t})$. A Cleanup decision point
renders as:
\begin{quote}\scriptsize
\texttt{Action menu:}\\
\texttt{A) eat apple @ (2, 11)}\\
\texttt{B) clean dirt @ (4, 5)}\\
\texttt{C) go to the river}\\
\texttt{D) go to the grass/orchard}\\
\textbf{Response:} \texttt{choice: B}
\end{quote}
(Schematic; a verbatim decision from the trained system appears in
\S\ref{app:example}.)

Departures from the strict options framework: our options are fixed and
symbolic (no option discovery, no termination learning); execution is
call-and-return ($\beta_z \in \{0,1\}$, never interrupted); menu
availability may consult state beyond the acting agent's view and is
computed centrally, while option selection and all learning remain
per-agent; and the library is sampled per training seed rather than given
as a single canonical set.

\subsubsection{How the option library is built.}
The pipeline has exactly one frontier-model step (Stage~1); Stages 2--4
are fixed algorithms with no model in the loop, and the guards are
counted from execution data, never suggested by a model or selected by a
human.

\paragraph{Stage 1: option-schema drafting (the only model step)}

Claude Opus 4.8 (Anthropic) is prompted with the game's source code and
asked to produce a typed option library in which every option carries a
four-part contract. The prompt is reproduced verbatim (Overcooked-AI,
Asymmetric Advantages; the Cleanup and Commons Harvest prompts follow the
same template with game-specific sections swapped):

\begin{quote}\scriptsize
\begin{verbatim}
You are designing an option library for an LLM
agent playing Overcooked-AI (layout:
asymmetric_advantages, onion-only recipes).
Produce a set of typed short-horizon options. For
each option give a 4-part contract:
  1. instantiation -- when does an instance appear
     in the menu (target existence only; NO
     state-conditional gating: do not encode game
     strategy or env legality rules -- those are
     unknown to you),
  2. executor -- symbolic expansion into primitive
     actions (BFS to an interact pose + INTERACT).
     The executor may fail explicitly only for
     reasons internal to expansion (no target, no
     path, nothing to drop).
  3. termination -- queue drained.
  4. effect predicate -- a post-condition on the
     state delta that is TRUE iff the execution
     did what the option name promises.
Do not add feasibility filters beyond what
expansion itself requires; the guard set will be
synthesized from execution data, not authored.
\end{verbatim}
\end{quote}

The generated schema implements the contracts as code. The raw executor is
deliberately environment-naive:

\begin{quote}\scriptsize
\begin{verbatim}
def raw_expand(inst, snap, feat, aid,
               blocked=None):
    # Checks ONLY: target given, path exists,
    # drop has an item.
    if inst.target is None:
        return Expansion(failure=
            FailureReason("TargetAbsent", inst.opt))
    if inst.opt == "drop_held" and held is None:
        return Expansion(failure=
            FailureReason("NothingHeld"))
    plan = shortest_interact_plan(inst.target, pos,
        orient, feat.walkable, blocked)
    if plan is None:
        return Expansion(failure=
            FailureReason("NoPath", str(inst.target)))
    return Expansion(actions=plan)
\end{verbatim}
\end{quote}

These explicit checks are \emph{not} authored gates in disguise: they are
algorithmic preconditions of expansion itself---the domain of the partial
function $\pi_z$ (path planning needs a target; a drop sequence needs an
item)---and the prompt whitelists exactly these reasons. They hide nothing
from the menu: the option still appears, the LLM can still select it, and
the failure is simply logged. An executor failure becomes a menu gate only
through Stage~3, by being counted in burn-in data like any other failure.
Environment-legality knowledge (a pot must be non-empty and idle to start
cooking) never appears in the executor; it is exactly what Stage~3 must
discover.

Effect predicates are written in \emph{delta} form---the repair loop below
explains why this is load-bearing---and shared-state effects are tied to a
private transition of the acting agent:

\begin{quote}\scriptsize
\begin{verbatim}
# acquire effects: delta form, not absolute
if inst.opt == "pick_dish":
    return post["held"] == "dish" \
       and pre["held"] != "dish"
# shared-state effect bound to the PRIVATE held
# transition, so a partner's placement is never
# attributed to this agent
if inst.opt == "place_in_pot":
    return pre["held"] in ("onion", "tomato") \
       and post["held"] is None \
       and pot_count_increased(pre, post)
\end{verbatim}
\end{quote}

\paragraph{Stage 2: uniform-over-menu burn-in (no LLM)}

A fixed harness runs episodes under a one-line policy---uniform random
over the instantiated, guardless menu:

\begin{quote}\scriptsize
\begin{verbatim}
inst = (policy(menu, snap, aid, rng)
        if policy else rng.choice(menu))
\end{verbatim}
\end{quote}

Each completed option execution emits one record: the option, its target,
the declared read-set features at initiation, and an outcome in
\{\texttt{success}, \texttt{effect\_fail}, \texttt{explicit:<reason>}\},
decided by the effect predicate or the executor's typed failure, never by
reward:

\begin{quote}\scriptsize
\begin{verbatim}
{"ep": 512, "step": 41, "aid": 0,
 "opt": "start_cooking", "target": [4, 2],
 "features": {"held": "onion",
              "tgt_pot_state": "full"},
 "outcome": "effect_fail"}
\end{verbatim}
\end{quote}

The declared read set contains only deterministic features
(\texttt{held}, \texttt{tgt\_pot\_state}, \texttt{tgt\_item}); transient
multi-agent features (e.g., partner adjacency) are logged but excluded
from guard eligibility a priori. On Overcooked Asymmetric Advantages,
$2{,}000$ solo plus $2{,}000$ two-agent episodes (${\sim}550$k decisions)
take about $70$ seconds on one CPU core; no GPU is used in Stages 2--4. A
control experiment replacing the uniform policy with a frozen pretrained
LLM collapses coverage---its narrow deterministic attractor never reaches
dish/plate/deliver states---yielding $2$ guards versus uniform's $36$ at
roughly $400\times$ the cost, which validates the uniform-burn-in choice.

\paragraph{Stage 3: guard synthesis and the acceptance certificate}

Synthesis runs in two channels over the burn-in log. \emph{Explicit}: a
typed failure reason observed at least five times becomes a guard
directly (\texttt{NoPath} $\to$ a live reachability check at menu build;
\texttt{NothingHeld} $\to$ drop requires a held item). \emph{Inferred}:
every 1- and 2-literal conjunction over the read set is scored and
adopted iff it covers at least five failures and \emph{zero} logged
successes:

\begin{quote}\scriptsize
\begin{verbatim}
for conj in cands:   # 1- and 2-literal conjs
    n_fail = sum(covers(conj, r) for r in fails)
    n_succ = sum(covers(conj, r) for r in succs)
    if n_fail >= MIN_SUPPORT and n_succ == 0:
        adopted.append(conj)   # certificate:
                               # hides ZERO successes
\end{verbatim}
\end{quote}

The purity requirement is the acceptance certificate, and it is what
keeps strategy out of the guard set: any conjunction that also covers a
success---including successes of temporally extended executions whose
precondition became true mid-walk---is rejected. Conjunctions subsumed by
an adopted single literal are dropped. Adopted guards are stored as data:

\begin{quote}\scriptsize
\begin{verbatim}
{"opt": "start_cooking", "channel": "inferred",
 "when": {"tgt_pot_state": "empty"},
 "support": 31887}
{"opt": "start_cooking", "channel": "inferred",
 "when": {"held": "dish"}, "support": 24240}
{"opt": "deliver", "channel": "explicit",
 "when": {"__reason__": "NoPath"},
 "support": 20884}
\end{verbatim}
\end{quote}

\paragraph{Stage 4: menu assembly}

Fixed code interprets the guard file: the menu instantiates each option
once per target of its class (never nearest-target only), evaluates every
adopted conjunction against each instance's read-set features, and hides
matches; explicit-channel guards are re-checked live (e.g.,
\texttt{NoPath} by running the path planner). Finally, menu letters are
shuffled per decision with an episode-seeded RNG: letter assignment is
arbitrary, and a fixed construction order would leak option identity
through position, which models exploit through a strong letter-A prior.
Training rollouts expand options with the same raw executor the guards
were mined against, so certified-valid temporally extended behaviors
(e.g., walking to a still-cooking pot that finishes en route) remain
executable.

\paragraph{The outer contract-repair loop}

Stages 2--4 form the inner, fully mechanical loop. The outer loop exists
because the \emph{contract itself}---the Stage-1 schema---can be
defective, and a defective contract corrupts the data the inner loop
learns from. The division of labor mirrors the rest of the pipeline: the
mechanical layers \emph{detect} a defect (it surfaces as an impossible
pattern in the burn-in log or the certificate), the frontier model
\emph{repairs} the schema, and Stages 2--3 rerun. Three defect families
arose in development; we walk through each as bug, symptom, and fix.

\emph{Defect 1: absolute-form effect predicates.} The effect of
\texttt{pick\_onion} was first written in absolute form: after execution,
\texttt{held\,=\,onion}. The bug: an agent that \emph{already} holds an
onion and executes \texttt{pick\_onion} performs a silent no-op---yet the
predicate is still true (it does hold an onion), so the vacuous execution
is logged as a \emph{success}. Symptom: these false successes flow into
the certificate and veto the correct guard---``hide \texttt{pick\_onion}
when \texttt{held\,=\,onion}'' is rejected for hiding logged
``successes'' that were never real. Fix: rewrite every acquire-type
effect in \emph{delta} form, \texttt{pre\,$\neq$\,X} $\wedge$
\texttt{post\,=\,X}. The no-op case now logs as \texttt{effect\_fail},
and the missing guard is adopted on the next synthesis pass.

\emph{Defect 2: shared-state attribution.} The effect of
\texttt{place\_in\_pot} was first ``pot count increases.'' In two-agent
burn-in, while agent~A is still walking to the pot, agent~B places an
onion; the pot count rises, and A's in-flight option is logged as a
success although A did nothing. Symptom: forged successes, again vetoing
correct guards through the certificate. Fix: bind every effect to a
\emph{private} transition of the acting agent (my held item goes
\texttt{onion} $\to$ \texttt{None}, \emph{and} the pot count rose); options
with no private component (\texttt{start\_cooking}) are synthesized from
solo-episode records only.

\emph{Defect 3: nearest-target instantiation.} Options were first
instantiated only for their \emph{nearest} target. In Asymmetric
Advantages the serving station nearest to one agent lies in the
unreachable half of the kitchen, so \texttt{deliver} always failed with
\texttt{NoPath} and effectively vanished from the menu---even though a
reachable station existed farther away. Symptom: an option whose explicit
failure rate is $100\%$ while the task it names is plainly achievable.
Fix: instantiate one menu entry \emph{per target} of each bounded class
(every pot, every station); guards then filter per instance.

\paragraph{Worked example: one decision on the trained system.}
\label{app:example}

A decision reproduced verbatim from a logged Overcooked Asymmetric
Advantages rollout of the trained Qwen3-4B checkpoint under the
auto-guard interface. The system message is fixed; menus are instantiated
per target and letters are shuffled per decision.

\begin{quote}\scriptsize
\begin{verbatim}
--- SYSTEM MESSAGE ---
You are an agent in a 2-player cooperative
Overcooked-AI kitchen.

Goal: place 3 onions in a pot, start the cook,
plate the soup with a dish, deliver to the serving
station. Each delivery scores +20 (shared with
partner).

The action menu only lists currently-feasible
actions. Pick one letter.

Output format (exactly two lines, nothing else):
state:  <empty | filling | cooking |
         ready_to_plate | delivering>
choice: <single capital letter>

--- USER MESSAGE ---
You are player 1. World position: (1, 1).
Facing: West. Holding: onion.
Partner (player 0): pos=(5, 2) facing=North
holding=onion.

Pots:
  - pot@(4, 2): 0/3 onions
      you: 4 | partner: 1  (partner closer)
  - pot@(4, 3): 0/3 onions
      you: 5 | partner: 2  (partner closer)
Counters: (empty)
Layout: asymmetric_advantages.   Step 4/150.

Memory:
- Last 5 macros decided: pick_onion
- Total reward last 100 steps: 0
- Partner: pos=(5, 2) facing=North holding=onion

Action menu:
A) place held item in pot @ (4, 2)
B) drop held item on counter @ (1, 0)
C) place held item in pot @ (4, 3)
D) drop held item on counter @ (2, 1)

Choose one. Reply with exactly two lines.

--- MODEL RESPONSE (verbatim) ---
state: cooking
choice: A
\end{verbatim}
\end{quote}

Three interface properties are visible. The pot options (A, C) are
instantiated once per target, and the counter-drop options (B, D) appear
un-gated: the auto pipeline adopts no usability gate on hand-offs, since
dropping while holding an item is a certified-valid execution. Letters
carry no meaning across decisions. And the response contract is two
plain lines; in the same logged episode, one of the eight recorded
decisions violated the format and failed to parse.

\subsection{Per-Agent LoRA and MAGRPO}
\label{sec:method-magrpo}

The collapse of \S\ref{sec:method-failures} has two distinct components, and
they call for different fixes. The first is a \emph{decoding-level prior
attractor}: the frozen base concentrates its output distribution on one
``safe'' action. On-policy policy gradient cannot repair this attractor---it
sharpens it, because only sampled actions receive gradient. The
option interface of \S\ref{sec:method-ip} attacks this component by
reshaping the action space. The second component arises in the multi-agent
setting: with a single shared adapter, all agents remain copies of one
policy---their rollouts explore only symmetric behavior, conflicting role
gradients average into the same weights, and the team settles into a joint
local optimum. This happens at the macro level even with the option
interface in place. Our fix simply makes
agents \emph{able to differ}: each agent $i$ owns a private LoRA adapter
$\theta_i$ on the shared frozen base $f$, so
$\pi^{i}_{\theta_i}(z^{i} \mid o^{i}) = f \circ \theta_i$, and each adapter
is updated only on its own agent's decisions. Differences introduced by rollout stochasticity can then
accumulate in private parameters rather than average away, letting the
team escape the symmetric attractor when the task rewards it.

We train these adapters with \emph{Multi-Agent Group Relative Policy Optimization} (MAGRPO); because each agent carries a private LoRA adapter on a shared frozen base, we refer to this per-agent variant as \emph{PA-MAGRPO}. At each training round we collect $K$ parallel joint rollouts of length $T$ under the current joint policy, using the option-menu interface from \S\ref{sec:method-ip} as the action interface. Each rollout is scored by a shared team return $R_k$ -- the team's task reward for that episode (total apples eaten in Cleanup and Commons Harvest, total soups delivered in Overcooked) -- identical for all agents in that rollout, and the group-relative advantage is computed by normalizing $R_k$ within the cohort.
The word ``per-agent'' refers to the parameterization and update, not to a separate reward: each agent has its own adapter, and each surrogate loss is evaluated only on that agent's decisions $\{(o^i_{k,t},z^i_{k,t})\}$. Specialization can therefore emerge because private adapters experience different observation--action histories under the same team advantage, rather than because cleaners and eaters receive different hand-designed rewards. Updates occur at option boundaries only; with episodic, undiscounted team returns this needs none of the semi-MDP discounting machinery. The full per-round procedure is given in Algorithm~\ref{alg:magrpo}.

Algorithm~\ref{alg:magrpo} states one PA-MAGRPO training round in full.
Each round collects $K$ joint rollouts under the current joint policy over
the option menu, scores every rollout by the shared team return, and
normalizes those returns within the cohort to obtain a single
group-relative advantage per rollout---the same value for every agent.
Each agent then takes PPO-clipped epochs over \emph{its own} option
decisions under that shared advantage, with a KL penalty toward the
reference policy, and only its private LoRA adapter is updated. Updates
occur at option boundaries only; with episodic, undiscounted team returns
this requires none of the semi-MDP discounting machinery.

\begin{algorithm}[h]
\caption{MAGRPO training round (per-agent LoRA over option-menu rollouts)}
\label{alg:magrpo}
\begin{algorithmic}[1]
\STATE \textbf{input:} base LLM $f$, per-agent adapters $\{\theta_{i}\}_{i=1}^{N}$, reference policy $\pi_{\text{ref}}$, option set $\mathcal{Z}$, rollouts $K$, episode length $T$, clip $\epsilon$, KL weight $\beta$, PPO epochs $E$.
\FOR{$k = 1, \dots, K$}
    \STATE Reset env; for $t = 1, \dots, T$, each agent $i$ samples an option $z^{i}_{k,t} \sim \pi^{i}_{\theta_{i}}(\cdot \mid o^{i}_{k,t})$ from the option menu $\mathcal{Z}^{i}(o^{i}_{k,t})$, the symbolic planner expands it into environment primitives, and the env returns $r_{k,t}$.
\ENDFOR
\STATE For each rollout $k$, accumulate the shared team return $R_k = \sum_t R(s_{k,t}, \mathbf{a}_{k,t})$.
\STATE Compute cohort statistics $\mu = \tfrac{1}{K}\!\sum_k R_k, \;\; \sigma = \mathrm{std}_k\,R_k,$ and the shared group-relative advantage
\[
\hat{A}_{k,t} \;=\; \frac{R_k - \mu}{\sigma}.
\]
\STATE For each agent $i$, take $E$ PPO-clipped epochs over its own decisions $\{(o^{i}_{k,t}, z^{i}_{k,t})\}$ to minimize
\begin{multline*}
\mathcal{L}^{i}(\theta_{i}) = - \mathbb{E}\!\Big[\min\!\big(\rho^{i}_{k,t}\hat{A}_{k,t}, \\
\mathrm{clip}(\rho^{i}_{k,t}, 1{-}\epsilon, 1{+}\epsilon)\, \hat{A}_{k,t}\big)\Big] + \beta\,\mathrm{KL}\!\left(\pi^{i}_{\theta_{i}}\,\|\, \pi_{\text{ref}}\right),
\end{multline*}
where $\rho^{i}_{k,t} = \pi^{i}_{\theta_{i}}(z^{i}_{k,t} \mid o^{i}_{k,t})\,/\, \pi^{i}_{\theta_{i,\text{old}}}(z^{i}_{k,t} \mid o^{i}_{k,t})$ is the per-decision importance ratio, and update $\theta_{i}$.
\STATE Save per-agent LoRA checkpoints
\end{algorithmic}
\end{algorithm}

\section{Experiments}
\label{sec:exp}

\subsection{Setup}
\label{sec:exp-setup}

\begin{table*}[t]
\centering
\caption{Cross-base PA-MAGRPO robustness across three games. Values are mean $\pm$ std over five independent training seeds. For each base model, the first row is the frozen base baseline (all zeros) and the second row is our method (per-agent LoRA + option menu + MAGRPO).}
\label{tab:cleanup-crossbase}
\begin{tabular}{lccc}
\toprule
Base model & Cleanup (apples) & Overcooked AA (deliveries) & Harvest \\
\midrule
\multirow{2}{*}{Qwen3-4B-Instruct-2507} & $0.00 \pm 0.00$ & $0.00 \pm 0.00$ & $0.00 \pm 0.00$ \\
 & $\mathbf{92.85 \pm 13.23}$ & $\mathbf{6.21 \pm 0.80}$ & $\mathbf{85.41 \pm 6.63}$ \\
\midrule
\multirow{2}{*}{Qwen3.5-2B} & $0.00 \pm 0.00$ & $0.00 \pm 0.00$ & $0.00 \pm 0.00$ \\
 & $\mathbf{105.69 \pm 50.71}$ & $\mathbf{7.31 \pm 1.14}$ & $\mathbf{97.82 \pm 6.82}$ \\
\midrule
\multirow{2}{*}{gemma-4-E4B-it} & $0.00 \pm 0.00$ & $0.00 \pm 0.00$ & $0.00 \pm 0.00$ \\
 & $\mathbf{60.25 \pm 39.40}$ & $\mathbf{7.08 \pm 1.15}$ & $\mathbf{99.84 \pm 15.76}$ \\
\midrule
\multirow{2}{*}{gemma-4-E2B-it} & $0.00 \pm 0.00$ & $0.00 \pm 0.00$ & $0.00 \pm 0.00$ \\
 & $\mathbf{68.62 \pm 41.59}$ & $\mathbf{6.60 \pm 0.84}$ & $\mathbf{115.52 \pm 12.02}$ \\
\bottomrule
\end{tabular}
\end{table*}

\begin{table*}[t]
\centering
\caption{Ablations on the fixed Qwen3-4B backbone. Values are mean $\pm$ std over 20 evaluation episodes; best scores are bold.}
\label{tab:ablation}
\resizebox{\linewidth}{!}{%
\begin{tabular}{lccc}
\toprule
& Cleanup (apples) & Overcooked AA (deliveries) & Harvest \\
\midrule
Frozen + native (Abl.~C)        & $0.00 \pm 0.00$ & $0.00 \pm 0.00$ & $0.00 \pm 0.00$ \\
Frozen + option menu (base)   & $0.00 \pm 0.00$ & $0.20 \pm 0.51$ & $76.63 \pm 18.24$ \\
Native + MAGRPO (Abl.~A)        & $0.00$$^\dagger$ & $0.00$$^\dagger$ & $0.00$$^\dagger$ \\
Compound-JSON + MAGRPO (Abl.~D) & $59.70 \pm 47.66$ & $0.00$$^\dagger$ & $91.74 \pm 16.49$ \\
Shared-LoRA + option menu (Abl.~B) & $0.10 \pm 0.45$ & $2.25 \pm 1.59$ & $90.44 \pm 18.19$ \\
\textbf{Full method (ours)}     & $\mathbf{93.25 \pm 16.84}$ & $\mathbf{6.80 \pm 1.89}$ & $\mathbf{95.68 \pm 15.63}$ \\
\bottomrule

\end{tabular}
}
\parbox{\linewidth}{\footnotesize $^\dagger$ Training collapsed at zero return for the entire training horizon; no positive checkpoint exists to evaluate.}

\end{table*}


\paragraph{Environments.}
We evaluate on 3 cooperative spatial games. \emph{Melting~Pot Cleanup}~\citep{leibo2017ssd, agapiou2022melting} (\texttt{clean\_up} substrate, 7 agents, 150-step episodes): agents share a river that must be cleaned of dirt, and an orchard that grows apples once dirt is below a threshold. The team-versus-individual tension creates a classic social dilemma. \emph{Overcooked-AI}~\citep{carroll2019overcooked} (2 agents, 400-step episodes): we use two separated-kitchen layouts. In Asymmetric Advantages (AA), the two cooks are separated but each can still reach all resources needed to complete soups; in Forced Coordination (FC), neither side can complete a soup alone, so agents must hand off onions across the counter. \emph{Commons~Harvest} (\texttt{commons\_harvest\_\allowbreak no\_thinning} substrate, 7 agents, 400-step episodes): agents eat apples on 6 patches for reward, but they must conserve, or the apples never regenerate on the patches. Full details of each environment follow.

\textbf{Melting Pot Cleanup} (\texttt{clean\_up} substrate; 7 agents;
150-step episodes). Agents share a river that accumulates dirt and an
orchard whose apples regrow only while river dirt stays below a threshold.
Eating apples yields individual reward that contributes to the team return;
cleaning yields none directly. The tension between harvesting now and
cleaning for future growth makes the game a sequential social dilemma:
sustained team reward requires some agents to forgo eating.

\textbf{Overcooked-AI} (2 agents; 400-step episodes; two separated-kitchen
layouts). Cooks assemble and deliver onion soups: fetch three onions into a
pot, cook, plate, and deliver. In \emph{Asymmetric Advantages} (AA) the two
cooks occupy separate kitchen halves but each half contains every resource
needed to complete a soup alone; coordination can help but is not required.
In \emph{Forced Coordination} (FC) neither half is self-sufficient---one
side has ingredient dispensers, the other the pots and serving
window---so nothing can be delivered without counter hand-offs. AA/FC thus
form a within-game control pair in which the same recipe faces optional
versus mandatory coordination.

\textbf{Commons Harvest} (\texttt{commons\_harvest\_\allowbreak no\_thinning} substrate;
7 agents; 400-step episodes). Agents eat apples growing on six shared
patches; an apple regrows only while at least one apple remains on its
patch, so exhausting a patch destroys it permanently. Team return favors
restraint over the 400-step horizon, but each individual bite is immediately
rewarded---a common-pool resource dilemma.

\textbf{Option-library sampling.} Frontier-model outputs cannot be
seeded, so the construction prompt (\S\ref{sec:method-ip}) is sampled
repeatedly and each training seed draws one candidate library at
random---our results are not contingent on any single hand-polished
library.

\paragraph{Models and implementation.}

Our main experiments train the recipe across four small backbones
(Qwen3-4B-Instruct-2507, Qwen3.5-2B, gemma-4-E4B-it, gemma-4-E2B-it;
``Qwen3-4B'' hereafter always denotes the first) with five seeds
each. The ablation suite then fixes that same Qwen3-4B backbone, so that
comparisons isolate the effect of the
action interface and fine-tuning design rather than differences between
backbones. For Cleanup and Commons Harvest we use LoRA rank 32, learning rate
$5\!\times\!10^{-5}$, $K\!=\!4$ rollouts per round, and sampling temperature
$0.7$. For Overcooked-AI we use LoRA rank 16, learning rate
$2\!\times\!10^{-5}$, $K\!=\!6$, sampling temperature $0.9$, and KL
coefficient $0.05$. All runs use PPO clip $\epsilon = 0.2$, two inner
epochs, advantage normalization, and gradient clipping at $1.0$. Training
runs on a single H100. These values were chosen from small pilot runs for
training stability under a fixed compute budget and then held fixed across
all methods compared within the same environment; no per-method tuning was
performed. Within a game, the full method and its
ablations share base model, rollout budget, optimizer settings,
temperature, checkpoint rule (highest training-time team return), and
evaluation protocol; ablations differ only in the removed component.
Final evaluation uses 20 episodes with seeds disjoint from training.

In the multi-seed, multi-backbone study, a training seed controls option-library
sampling, LoRA initialization, environment initial states, and minibatch
order, so each seed is an independent end-to-end run. The ablation suite
(Table~\ref{tab:ablation}) uses a single training seed on the fixed
Qwen3-4B backbone to keep the six-cell design affordable; the cross-base
study (Table~\ref{tab:cleanup-crossbase}) tests whether the pattern
persists across seeds and model families.
\paragraph{What the training seed controls.}
A training seed fixes every stochastic component of one end-to-end run,
and its influence enters through four channels. (i)~\emph{Interface}: the
seed selects one of the retained option-library designs, so different
seeds train against different option sets, menu phrasings, and per-decision
letter shuffles---a different effective action space.
(ii)~\emph{Initialization}: per-agent LoRA weights are initialized from
the seed, which determines the initial symmetry-breaking among the $N$
otherwise identical agents and hence which agent drifts toward which role.
(iii)~\emph{Rollout stochasticity}: environment initial states and the
temperature-based sampling of options during the $K$ rollouts are seeded;
because MAGRPO advantages are group-relative, whichever behavior first
scores above the cohort mean is amplified. (iv)~\emph{Optimization order}:
minibatch shuffling within the PPO epochs. Channels (ii)--(iv) make each
seed an independent optimization trajectory even under a fixed interface;
channel (i) additionally varies the interface itself. The per-seed
standard deviations in the cross-base table of the main paper therefore
reflect the variability of the entire pipeline, not evaluation noise
alone; evaluation episodes use a fixed held-out seed set and are
unaffected by the training seed.

\paragraph{Evaluation protocol and statistics.}
Every cell of the ablation table in the main paper is a 20-episode
evaluation on the fixed seed set $\{10{,}000, 10{,}001, \dots, 10{,}019\}$,
disjoint from the training seed range, so all rows are comparable
seed-for-seed. Headline cells satisfy 95\% percentile bootstrap confidence
intervals that exclude zero (ours, Cleanup: $93.25\,[86.05,\,100.85]$;
ours, Overcooked AA: $6.80\,[5.95,\,7.60]$; $10{,}000$ resamples each).
All pairwise differences between the full method and the frozen baselines
are significant at $p < 10^{-7}$ under a two-sided Mann--Whitney $U$ test. Because the ablation suite uses a single training seed, those intervals quantify evaluation-side variability only; the cross-base study (Table~\ref{tab:cleanup-crossbase}) and both behavioral figures aggregate five independent training seeds.

\paragraph{Ablations.}
The ablation suite was run on a slightly earlier interface version
whose guards were hand-selected rather than synthesized; since the
ablations isolate the value of the menu and of per-agent capacity---not
how guards are obtained---we report them unchanged. We compare six cells drawn from the design space \{\textbf{abstraction}: option menu / compound-JSON / native actions\} \(\times\) \{\textbf{capacity}: per-agent LoRA / single shared LoRA / no training\}.
\begin{itemize}
\item \textbf{Full method (ours)} — option menu + per-agent LoRA + MAGRPO.
\item \textbf{Ablation A (native + MAGRPO)} — drop the abstraction, keep per-agent LoRA, train with MAGRPO over raw motor actions.
\item \textbf{Ablation B (option menu + shared LoRA + MAGRPO)} — keep the abstraction, replace seven adapters with one shared adapter; all agents back-propagate into the same LoRA.
\item \textbf{Ablation C (frozen, native)} — no training, no abstraction; sanity check that the base model alone cannot play.
\item \textbf{Ablation D (compound-JSON + per-agent LoRA + MAGRPO)}
replaces the menu with free structured generation over the same family of
compound actions: the LLM emits a JSON object specifying an action type
and arguments, executed when valid. This keeps macro-level action
semantics but removes feasibility filtering and discrete menu
selection---separating the value of macro actions from the value of the
state-conditioned menu.
\end{itemize}

\subsection{Results}
\label{sec:exp-results}
\paragraph{Main result: the recipe works across backbones and seeds.}
Across four small backbones $\times$ five independent training seeds
(Table~\ref{tab:cleanup-crossbase}), the frozen base collects zero in
every row of every game, while our full method obtains positive return
for every backbone: Cleanup $60.25$--$105.69$, Overcooked Asymmetric
Advantages $6.21$--$7.31$, Commons Harvest $85.41$--$115.52$. Gain
magnitudes vary with backbone and seed---interface design and
initialization still affect optimization---but the qualitative pattern is
stable: the frozen base never plays; menu-based per-agent fine-tuning
consistently produces non-trivial behavior.

\paragraph{Headline numbers on the fixed backbone.}
The full method delivers large gains over the frozen base on every game we evaluate (Table~\ref{tab:ablation}). On Melting Pot Cleanup the team rises from $0.00\!\pm\!0.00$ apples (frozen base with raw motor actions) to $93.25\!\pm\!16.84$ apples per 150-step episode. On Overcooked-AI Asymmetric Advantages deliveries rise from $0.20\!\pm\!0.51$ to $6.80\!\pm\!1.89$ per 400-step episode -- a $34\times$ improvement. On Commons Harvest the team reaches $95.68\!\pm\!15.63$ apples. Every ``ours'' cell is statistically separated from its frozen and zero-shot cells.
\paragraph{Removing the option menu is catastrophic (Ablation~A).}
Training the same per-agent LoRAs with MAGRPO directly on raw motor actions, with the option menu switched off, collapses every cell to zero on all three games and never produces a positive checkpoint. This confirms the statement of \S\ref{sec:method-magrpo}: at the raw-action timescale the 4B base rarely emits a coherent multi-step trajectory under random exploration, so cohort returns are nearly always zero, the GRPO group standard deviation collapses, and updates carry no useful signal---gradients cannot repair what sampling never reaches.
\paragraph{A single shared LoRA cannot hold role specialization (Ablation~B).} Keeping the menu but sharing one adapter recovers the reward signal, yet the joint policy never stabilizes: Cleanup averages $0.10\!\pm\!0.45$ apples at its best checkpoint (isolated rounds spike to ${\sim}55$, then revert---the adapter can briefly express one role but cannot hold cleaner \emph{and} eater at once), and Overcooked reaches $2.25\!\pm\!1.59$ deliveries, about a third of the full method. The symmetric Commons Harvest is forgiving ($90.44\!\pm\!18.19$), so the shared-vs.-per-agent gap is specifically about role differentiation, not capacity---confirming the statement of \S\ref{sec:method-magrpo} that a shared adapter re-creates the collapse at the macro level even with the option interface in place.
\paragraph{The frozen base cannot play (Ablation~C, and the ``Frozen + option menu'' baseline).} With no training and no abstraction the 4B base scores zero on every game, as expected. Adding the option menu to the frozen base (no MAGRPO) lifts Harvest to $76.63\!\pm\!18.24$ but leaves Cleanup at $0.00$ and Overcooked at $0.20\!\pm\!0.51$. This isolates the contributions cleanly: the abstraction alone is sufficient to play the easy, symmetric substrate (Harvest), but the harder cooperative games (Cleanup, Overcooked) additionally require MAGRPO fine-tuning of per-agent LoRAs.
\paragraph{Compound actions without the menu filter are unstable (Ablation~D).} Free-form compound-JSON generation---same macro semantics, no feasibility filter---partially trains on Cleanup ($59.70\!\pm\!47.66$, vs.\ ours $93.25\!\pm\!16.84$; the standard deviation is nearly as large as the mean) but never escapes zero on Overcooked; symmetric Harvest is again forgiving ($91.74\!\pm\!16.49$). Without the filter the model frequently emits compound actions that are illegal or no-ops in the current state (\S\ref{sec:exp-trajectory}). Across the four ablations, neither ingredient is redundant, as claimed in the intro: the menu is what produces \emph{any} non-zero return on the harder games, and per-agent LoRA is what lets the cohort \emph{hold} a heterogeneous joint policy once it exists. Harvest needs neither roles nor workflows to collect an apple, so it performs decently even untuned; what matters there is conservation, examined in \S\ref{sec:exp-emergent}.


\subsection{Trajectory-level analysis: why and how it works}
\label{sec:exp-trajectory}

To understand the gap between our method and the failed ablations, we record full per-step trajectories.
\paragraph{(1) The abstraction collapses the effective action space.}
In the native-action cells (A, C) every environment step triggers an LLM call (call ratio $1.00$). In the option-menu cells (B, ours) the ratio drops to ${\sim}0.40$ on Cleanup and Commons Harvest and $0.71$ on Overcooked: each LLM commitment spans $2$--$3$ env steps, so the MAGRPO advantage propagates over a $2$--$3\times$ shorter decision sequence.
\paragraph{(2) Native-action MAGRPO collapses to a single letter.}
The frozen base and native-action MAGRPO cells spend nearly all decisions on one motor action ($\texttt{FIRE\_CLEAN}$: $95.8\%$ frozen, $98.4\%$ after MAGRPO on Cleanup; $\texttt{EAST}$: $58.8\%$/$60.2\%$ on Overcooked). RL over raw letters does not learn a better policy---it sharpens an already-degenerate mode, exactly as claimed in \S\ref{sec:method-magrpo}: only sampled actions receive gradient, so on-policy RL reinforces the prior attractor instead of repairing it.

\paragraph{(3) Shared LoRA trains, but symmetry caps it.}
The shared-LoRA cell does escape zero reward, but a single adapter cannot
hold two roles at once: every agent converges to the same macro
distribution. On Cleanup its training curve shows brief reward spikes when
the one policy happens to express a useful role (everyone cleans, or
everyone eats), followed by reversion as the complementary work goes
undone---a symmetric policy that looks locally sensible per agent yet
makes an inefficient team, exactly the mechanism per-agent adapters remove.

\begin{figure*}[t]
\centering
\includegraphics[width=0.82\linewidth]{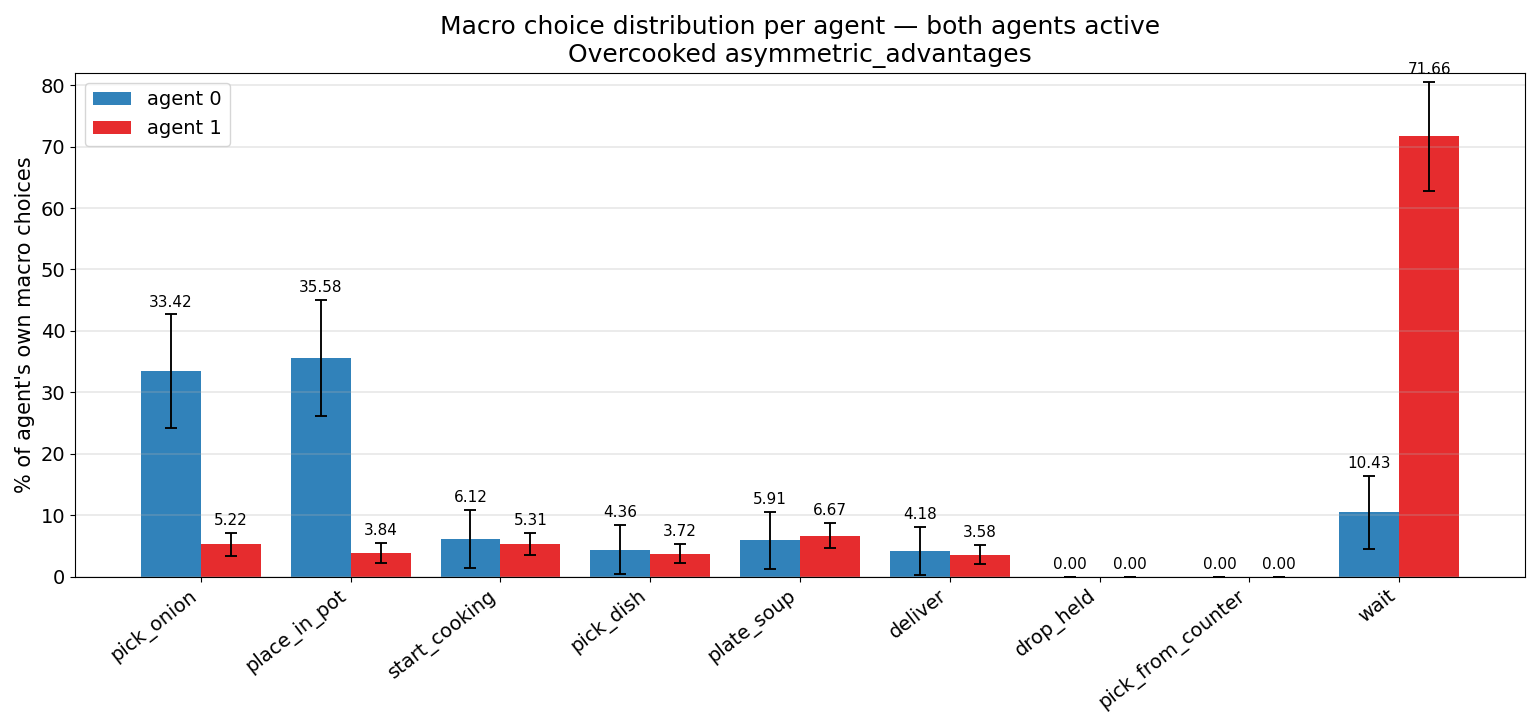}
\caption{Per-agent macro choices on Overcooked AA. Agent~0 concentrates on onion loading ($\approx69\%$); agent~1 \texttt{wait}s $\approx72\%$ of the time; counter operations are zero for both---no hand-off, a one-worker/one-idler split.}
\label{fig:oc-behavior-hist}
\end{figure*}

\begin{figure*}[t]
\centering
\includegraphics[width=0.82\linewidth]{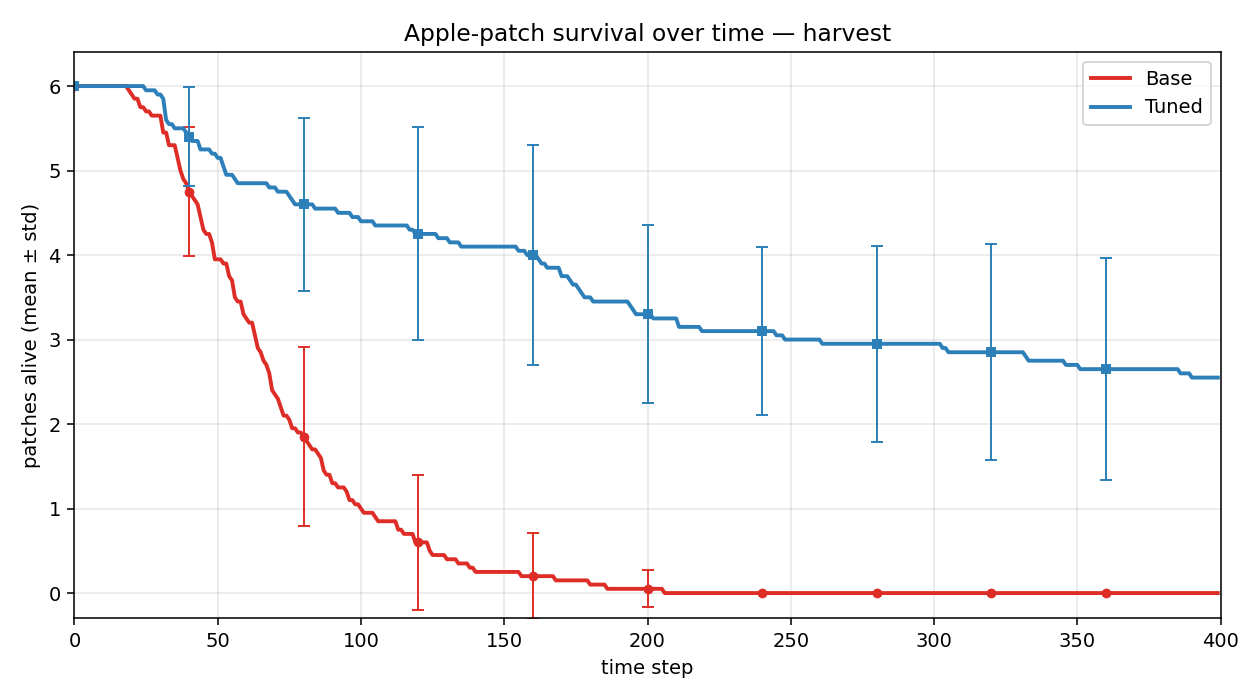}
\caption{Commons Harvest patch survival. The frozen base depletes all patches by ${\sim}$step~200; the tuned checkpoint keeps ${\sim}2.5$ of $6$ alive through step~400---restraint despite a nearly flat reward.}
\label{fig:harvest-patch-survival}
\end{figure*}

\begin{figure*}[t]
\centering
\includegraphics[width=0.98\linewidth]{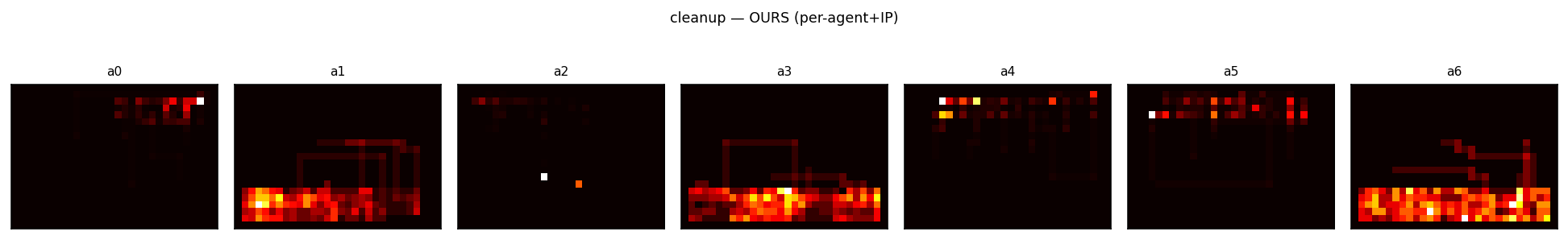}
\caption{Per-agent visit heatmaps for Cleanup (ours). With no role labels, $\text{a}_1,\text{a}_3,\text{a}_6$ specialize as eaters (orchard rows) and $\text{a}_0,\text{a}_4,\text{a}_5$ as cleaners (river shore); $\text{a}_2$ is near-stationary.}
\label{fig:heatmap}
\end{figure*}

\subsection{Emergent behavior: reward and cooperation decouple}
\label{sec:exp-emergent}

Beyond \emph{why} the method trains, the trajectories reveal \emph{what} the agents do: under one fixed recipe, genuine role specialization on Cleanup, no coordination on Overcooked, and resource restraint on Commons Harvest. 
\paragraph{Cleanup: emergent role specialization.}
The seven Cleanup adapters of our method partition cleanly into ``cleaner'' and ``eater'' roles without any role label: per-agent apple counts split into three eaters (${\sim}57$--$60$ apples each), three cleaners (all $0$), and one near-stationary helper, and the visit heatmaps (Fig.~\ref{fig:heatmap}) confirm the spatial split---eaters on the orchard rows, cleaners on the river shore. Shared LoRA (Abl.~B) cannot reproduce this: its per-agent breakdown is essentially uniform. Strikingly, compound-JSON with per-agent LoRA (Abl.~D) \emph{also} fails to specialize---all seven adapters eat apples, and the identity of the dominant eater shifts from episode to episode. Per-agent capacity is therefore necessary but not sufficient: the constrained pick of the option menu is what lets each adapter commit to a distinct typed-option distribution that gradient descent can sharpen agent-by-agent, whereas free-form $(x,y)$ generation diffuses the gradient and the partition collapses.

\paragraph{Overcooked: high reward without coordination.}
On Overcooked-AI Asymmetric Advantages (Table~\ref{tab:ablation}), each agent can reach every resource, and the trained team reaches high reward without coordinating. The per-agent macro histogram (Fig.~\ref{fig:oc-behavior-hist}) shows two signatures: no item hand-off (counter operations are $0\%$ for both agents), and a lopsided workload---agent~0 spends ${\sim}69\%$ of choices loading onions while covering every downstream stage, whereas agent~1 \texttt{wait}s on $72\%$ of its choices. Run alone in the same layout, both agents still deliver ($4$ and $2$), so each has independently learned the full task: Asymmetric Advantages is solo-solvable, and reward here overstates cooperation. Forced Coordination removes the solo solution: neither cook can complete a soup alone. There the frozen base scores $0.00$ while the trained checkpoint reaches $7.57 \pm 1.16$ deliveries, and the same solo test flips---Forced-Coordination agents placed alone deliver $0$ each. Their positive team deliveries are behavior that only functions with a partner---direct evidence of emerged coordination.

\paragraph{Commons Harvest: restraint without reward.}
Because apple count does not reveal whether the shared patches survive, we track patch survival in a separate 400-step evaluation. Apple reward moves modestly (frozen $76.73 \pm 16.29$ vs.\ tuned $98.29 \pm 14.52$), but patch survival changes sharply (Fig.~\ref{fig:harvest-patch-survival}): the frozen base depletes all six patches by roughly step~$200$, while the tuned checkpoint keeps about $2.5$ alive through step~$400$. The mechanism is restraint: roughly two agents harvest while the rest wait, trading immediate consumption for persistence the reward curve barely registers.



\section{Conclusion and Limitations}
\label{sec:conclusion}

Casting small (2--4B) LLMs as policies over a frontier-drafted library of
symbolic options and training a private LoRA adapter per agent with
PA-MAGRPO makes them trainable as cooperative spatial policies (Cleanup
apples $0.00\!\to\!92.85$; Overcooked deliveries $0.00\!\to\!6.21$). Used as a fixed instrument, the same recipe reveals the decoupling:
high reward coincides with real specialization on Cleanup but with a
one-worker/one-idler team on Overcooked, nearly flat reward hides real
restraint on Commons Harvest, and coordination appears when---and only
when---the layout requires it. Team reward alone cannot certify cooperation; behavioral metrics must sit alongside it.
\paragraph{Limitations.}
Our results cover small ($\leq$4B) LLMs and three spatial games;
transfer to larger scales or non-spatial tasks is untested, the option
library is per-game, and we lack a principled way to choose an option
space. The decoupling finding is behavioral evidence, not a general theory:
the cross-game pattern replicates over five seeds and four backbones,
but the \emph{specific} role assignment is checked on one seed, and the
Forced-Coordination control assumes a non-solo-solvable layout.

\section*{Generative AI Disclosure}
Beyond the frontier model used inside the method itself (Claude Opus 4.8;
\S\ref{sec:method-ip}), large language models
(Claude Opus 5 and Claude Fable 5, Anthropic) were used to edit and revise
the manuscript draft. The authors remain fully responsible for all
content, and all references have been verified against their original
sources.

\bibliographystyle{aaai2027}
\bibliography{references}

\begin{thebibliography}{24}
\providecommand{\natexlab}[1]{#1}

\bibitem[{Agapiou et~al.(2022)Agapiou, Vezhnevets, Du{\'e}{\~n}ez-Guzm{\'a}n,
  Matyas, Mao, Sunehag, K{\"o}ster et~al.}]{agapiou2022melting}
Agapiou, J.~P.; Vezhnevets, A.~S.; Du{\'e}{\~n}ez-Guzm{\'a}n, E.~A.; Matyas,
  J.; Mao, Y.; Sunehag, P.; K{\"o}ster, R.; et~al. 2022.
\newblock Melting Pot 2.0.
\newblock \emph{arXiv preprint arXiv:2211.13746}.

\bibitem[{Ahn et~al.(2022)Ahn, Brohan, Brown et~al.}]{ahn2022saycan}
Ahn, M.; Brohan, A.; Brown, N.; et~al. 2022.
\newblock Do As I Can, Not As I Say: Grounding Language in Robotic Affordances.
\newblock In \emph{Conference on Robot Learning (CoRL)}.

\bibitem[{Amato, Konidaris, and Kaelbling(2014)}]{amato2014macdec}
Amato, C.; Konidaris, G.; and Kaelbling, L.~P. 2014.
\newblock Planning with Macro-Actions in Decentralized {POMDPs}.
\newblock In \emph{Proceedings of the 13th International Conference on
  Autonomous Agents and Multiagent Systems (AAMAS)}.

\bibitem[{Biswas et~al.(2026)Biswas, Palod, Bhambri, and
  Kambhampati}]{biswas2025interdependence}
Biswas, U.; Palod, V.; Bhambri, S.; and Kambhampati, S. 2026.
\newblock Who Is Helping Whom? Analyzing Inter-Dependencies to Evaluate
  Cooperation in Human-AI Teaming.
\newblock In \emph{Proceedings of the AAAI Conference on Artificial
  Intelligence}, volume~40, 17347--17356.

\bibitem[{Carroll et~al.(2019)Carroll, Shah, Ho, Griffiths, Seshia, Abbeel, and
  Dragan}]{carroll2019overcooked}
Carroll, M.; Shah, R.; Ho, M.~K.; Griffiths, T.~L.; Seshia, S.~A.; Abbeel, P.;
  and Dragan, A. 2019.
\newblock On the Utility of Learning about Humans for Human-{AI} Coordination.
\newblock In \emph{Advances in Neural Information Processing Systems
  (NeurIPS)}.

\bibitem[{Gallego(2026)}]{gallego2026cooperation}
Gallego, V. 2026.
\newblock Cooperation and Exploitation in {LLM} Policy Synthesis for Sequential
  Social Dilemmas.
\newblock \emph{arXiv preprint arXiv:2603.19453}.

\bibitem[{Hu et~al.(2022)Hu, Shen, Wallis, Allen-Zhu, Li, Wang, Wang, and
  Chen}]{LoraHu2021}
Hu, E.~J.; Shen, Y.; Wallis, P.; Allen-Zhu, Z.; Li, Y.; Wang, S.; Wang, L.; and
  Chen, W. 2022.
\newblock {LoRA}: Low-Rank Adaptation of Large Language Models.
\newblock In \emph{International Conference on Learning Representations
  (ICLR)}.

\bibitem[{Hua et~al.(2025)Hua, Chen, Wang, Li, Wang, and
  Luo}]{hua2025shapleycoop}
Hua, Y.; Chen, H.; Wang, S.; Li, W.; Wang, X.; and Luo, J. 2025.
\newblock Shapley-Coop: Credit Assignment for Emergent Cooperation in
  Self-Interested LLM Agents.
\newblock In \emph{Advances in Neural Information Processing Systems
  (NeurIPS)}.
\newblock ArXiv:2506.07388.

\bibitem[{Huang et~al.(2022)Huang, Xia, Xiao et~al.}]{huang2022innermonologue}
Huang, W.; Xia, F.; Xiao, T.; et~al. 2022.
\newblock Inner Monologue: Embodied Reasoning through Planning with Language
  Models.
\newblock In \emph{Conference on Robot Learning (CoRL)}.

\bibitem[{Hughes et~al.(2018)Hughes, Leibo, Phillips, Tuyls,
  Du{\'e}{\~n}ez-Guzm{\'a}n, Casta{\~n}eda, Dunning, Zhu, McKee, Koster
  et~al.}]{hughes2018inequity}
Hughes, E.; Leibo, J.~Z.; Phillips, M.; Tuyls, K.; Du{\'e}{\~n}ez-Guzm{\'a}n,
  E.~A.; Casta{\~n}eda, A.~G.; Dunning, I.; Zhu, T.; McKee, K.~R.; Koster, R.;
  et~al. 2018.
\newblock Inequity Aversion Improves Cooperation in Intertemporal Social
  Dilemmas.
\newblock In \emph{Advances in Neural Information Processing Systems
  (NeurIPS)}.

\bibitem[{Lee, Cho, and Choi(2026)}]{lee2026mapcoderlite}
Lee, W.; Cho, J.; and Choi, J. 2026.
\newblock MapCoder-Lite: Distilling Multi-Agent Coding into a Single Small LLM.
\newblock In \emph{Findings of the Association for Computational Linguistics:
  EACL}.
\newblock ArXiv:2509.17489.

\bibitem[{Leibo et~al.(2017)Leibo, Zambaldi, Lanctot, Marecki, and
  Graepel}]{leibo2017ssd}
Leibo, J.~Z.; Zambaldi, V.; Lanctot, M.; Marecki, J.; and Graepel, T. 2017.
\newblock Multi-Agent Reinforcement Learning in Sequential Social Dilemmas.
\newblock In \emph{Proceedings of the 16th International Conference on
  Autonomous Agents and Multiagent Systems (AAMAS)}.

\bibitem[{Liang et~al.(2023)Liang, Huang, Xia, Xu, Hausman, Ichter, Florence,
  and Zeng}]{liang2023codeaspolicies}
Liang, J.; Huang, W.; Xia, F.; Xu, P.; Hausman, K.; Ichter, B.; Florence, P.;
  and Zeng, A. 2023.
\newblock Code as Policies: Language Model Programs for Embodied Control.
\newblock In \emph{IEEE International Conference on Robotics and Automation
  (ICRA)}.

\bibitem[{Liu et~al.(2026)Liu, Liang, Lyu, and Amato}]{liu2025magrpo}
Liu, S.; Liang, Z.; Lyu, X.; and Amato, C. 2026.
\newblock LLM Collaboration with Multi-Agent Reinforcement Learning.
\newblock In \emph{Proceedings of the AAAI Conference on Artificial
  Intelligence}.
\newblock ArXiv:2508.04652.

\bibitem[{Ma et~al.(2024)Ma, Hu, Pu, Liu, Ai, Liang, and Chen}]{Cory}
Ma, H.; Hu, T.; Pu, Z.; Liu, B.; Ai, X.; Liang, Y.; and Chen, M. 2024.
\newblock Coevolving with the other you: Fine-tuning llm with sequential
  cooperative multi-agent reinforcement learning.
\newblock \emph{Advances in Neural Information Processing Systems}, 37:
  15497--15525.

\bibitem[{Mosquera et~al.(2025)Mosquera, Pinzon, Rios, Fonseca, Giraldo,
  Quijano, and Manrique}]{mosquera2024meltingpot}
Mosquera, M.; Pinzon, J.~S.; Rios, M.; Fonseca, Y.; Giraldo, L.~F.; Quijano,
  N.; and Manrique, R. 2025.
\newblock Can {LLM-Augmented} Autonomous Agents Cooperate? An Evaluation of
  Their Cooperative Capabilities through {Melting Pot}.
\newblock \emph{IEEE Transactions on Artificial Intelligence}.

\bibitem[{P{\'e}rolat et~al.(2017)P{\'e}rolat, Leibo, Zambaldi, Beattie, Tuyls,
  and Graepel}]{perolat2017commonpool}
P{\'e}rolat, J.; Leibo, J.~Z.; Zambaldi, V.; Beattie, C.; Tuyls, K.; and
  Graepel, T. 2017.
\newblock A Multi-Agent Reinforcement Learning Model of Common-Pool Resource
  Appropriation.
\newblock In \emph{Advances in Neural Information Processing Systems
  (NeurIPS)}.

\bibitem[{Piche et~al.(2025)Piche, Muqeeth, Aghajohari, Duque, Noukhovitch, and
  Courville}]{piche2025robust}
Piche, D.; Muqeeth, M.; Aghajohari, M.; Duque, J.; Noukhovitch, M.; and
  Courville, A. 2025.
\newblock Learning Robust Social Strategies with Large Language Models.
\newblock \emph{arXiv preprint arXiv:2511.19405}.

\bibitem[{Shao et~al.(2024)Shao, Wang, Zhu, Xu, Song, Bi, Zhang, Zhang, Li, Wu
  et~al.}]{grpo}
Shao, Z.; Wang, P.; Zhu, Q.; Xu, R.; Song, J.; Bi, X.; Zhang, H.; Zhang, M.;
  Li, Y.; Wu, Y.; et~al. 2024.
\newblock Deepseekmath: Pushing the limits of mathematical reasoning in open
  language models.
\newblock \emph{arXiv preprint arXiv:2402.03300}.

\bibitem[{Sutton, Precup, and Singh(1999)}]{sutton1999options}
Sutton, R.~S.; Precup, D.; and Singh, S. 1999.
\newblock Between {MDPs} and Semi-{MDPs}: A Framework for Temporal Abstraction
  in Reinforcement Learning.
\newblock \emph{Artificial Intelligence}, 112(1-2): 181--211.

\bibitem[{Valmeekam et~al.(2023)Valmeekam, Marquez, Sreedharan, and
  Kambhampati}]{valmeekam2023planning}
Valmeekam, K.; Marquez, M.; Sreedharan, S.; and Kambhampati, S. 2023.
\newblock On the Planning Abilities of Large Language Models -- A Critical
  Investigation.
\newblock In \emph{Advances in Neural Information Processing Systems
  (NeurIPS)}.
\newblock ArXiv:2305.15771.

\bibitem[{Xiao, Hoffman, and Amato(2020)}]{xiao2020macro}
Xiao, Y.; Hoffman, J.; and Amato, C. 2020.
\newblock Macro-Action-Based Deep Multi-Agent Reinforcement Learning.
\newblock In \emph{Proceedings of the 3rd Conference on Robot Learning (CoRL)},
  volume 100 of \emph{Proceedings of Machine Learning Research}, 1146--1161.

\bibitem[{Zhang, Kapoor, and Sun(2025)}]{zhang2025lorasa}
Zhang, B.; Kapoor, A.; and Sun, M. 2025.
\newblock Low-Rank Agent-Specific Adaptation (LoRASA) for Multi-Agent Policy
  Learning.
\newblock \emph{arXiv preprint arXiv:2502.05573}.

\bibitem[{Zhuang et~al.(2025)Zhuang, Shen, Zhang, Chen, and Miao}]{YoloMarl}
Zhuang, Y.; Shen, Y.; Zhang, Z.; Chen, Y.; and Miao, F. 2025.
\newblock {YOLO}-{MARL}: You Only {LLM} Once for Multi-agent Reinforcement
  Learning.
\newblock In \emph{IEEE/RSJ International Conference on Intelligent Robots and
  Systems (IROS)}.

\end{thebibliography}

\end{document}